\documentstyle[psfig]{elsart}

\newcommand{\be}{\begin{equation}}
\newcommand{\ee}{\end{equation}}
\newcommand{\bea}{\begin{eqnarray}}
\newcommand{\eea}{\end{eqnarray}}

\newcommand{\avg}[1]{\langle{#1}\rangle}

\newcommand{\BE}{\begin{eqnarray}}
\newcommand{\EE}{\end{eqnarray}}
\newcommand{\BEn}{\begin{eqnarray*}}
\newcommand{\EEn}{\end{eqnarray*}}
\newcommand{\barr}{\begin{array}}
\newcommand{\earr}{\end{array}}
\newcommand{\bit}{\begin{itemize}}      
\newcommand{\eit}{\end{itemize}}
\newcommand{\bc}{\begin{center}}
\newcommand{\ec}{\end{center}}
\newcommand{\ben}{\begin{enumerate}}    
\newcommand{\een}{\end{enumerate}}

\begin{document}

%@twocolumnfalse\endcsname
\begin{frontmatter}
\title{Analyzing and modelling 1+1$d$ markets}
\author{Damien Challet and Robin Stinchcombe}
\address{Theoretical Physics, 1 Keble Road, Oxford OX1 3NP, United Kingdom}
\date{\today}
\maketitle

\begin{abstract}
We report a statistical analysis of the Island ECN (NASDAQ) order book. We determine the static and dynamic properties of this system, and then analyze them from a physicist's viewpoint using an equivalent particle system obtained by treating orders as massive particles and price as position. We identify the fundamental dynamical processes, test existing particles models of such markets against our findings, and introduce a new model of limit order markets.
\end{abstract}

\end{frontmatter}

\section{Introduction}
Physicists can contribute much to the study of interacting humans, because of their familiarity with cooperative phenomena. Collective economic behavior is an area in which physicists have recently become acutely interested, probably because of the availability of a huge amount of free electronic data related to economy. In particular, physicists' attention \cite{web} has focused on financial markets \cite{Mandelbrot,Bouchaud,MantegnaStanley}. 

Most of the time, the only information one can obtain about a particular stock is the price and volume of the last transaction, and the current price and number of shares of the best buying offer (also called the highest bid), and, of course, the  price and number of shares of the best selling offer (lowest ask).

However, in electronic limit order book markets\footnote{Like for instance the NASDAQ, the {\em Bourse de Paris}, the Swiss Market (SMI), the Australian Stock Exchange (ASX), \dots}, more data are available, either publicly or by request: one knows not only the highest bid (buy) and the lowest ask (sell) orders, but also several lower bid and higher ask orders. This obviously gives more information than only the last price and volume, but it is several fold more difficult to analyze: in addition to the last price and volume time series, one has to deal with two time-evolving unidimensional order distributions.

There is growing literature on the analysis of limit order book markets, particularly in Economics \cite{Eco}. In Econophysics however, this area is almost a {\em terra incognita}, except for a very recent work \cite{MaslovMills}, where the five best prices on both bid and ask sides of the global NASDAQ book  are analyzed.

Here, we analyze data obtained from a sub-part of the NASDAQ, namely the Island Electronic Communication Network (ECN) (www.island.com), which displays in real time the 15 best orders of both types\footnote{Note that two orders of the same kind may have the same price.}. While previous studies have mainly addressed static properties of such markets, we provide a dynamic analysis that is able to identify some fundamental properties of such markets. In this way, by analyzing the available data from an appropriate point of view, we are able to arrive at the necessary ingredients for a minimal market model.
\section{Definition of the market}
There is one order book per stock in a given ECN (but there are several ECNs that are part of the NASDAQ); it reflects the current state of investors public wishes by listing what quantity of shares they are willing to buy or sell and at which price.
 An order $k$ is characterized by three intrisic quantities:
\bit
\item its type, bid (buy) or ask (sell); for each order $k$ this can be represented by a binary variable  $a_k\in\{B,A\}\equiv\{+1,-1\}$;
\item its price $p_k$, which is discretized;
\item its size $m_k$, in shares.
\eit

In addition, the owner of an order can specify a timeout, time interval after which the order is automatically removed if the order has not been filled. The life of this order $(a_k,p_k,m_k)$ begins at time $t_k^b$ when it is placed at price $p_k$. Possible prices are discretized, and the elementary step is called the tick --- in the Island ECN, a tick is equal to \$~1/256. It appears on the order book until time $t_k^e$ when it is removed\footnote{This may be due to a cancellation by its owner, or by its automatic removal, once its time-out is reached.}, or when a new order $(-a_k,p_k,m')$ of the opposite type and of at least the same size ($m'\ge m_k$) is placed at the same price, leading to a transaction\footnote{This is called a price-taking behavior.}; order k's lifetime is then $\tau_k=t_k^e-t_k^b$. Note that about $9\pm 1$\% of all orders happen to be partly filled at some point, that is, since $t_k^b$, an opposite order of a smaller size has been placed at price $p_k$. Therefore, the size of an order is not a conserved quantity. There are of course transaction costs: Island ECN charges \$~0.0025 per share for a price taking orders, and \$~0.0015 per share for the other orders\footnote{thus rewarding patient investors.}, but removing an order is free.

\section{The data}
We collected high frequency\footnote{Every one or two seconds on average.} data from the Island ECN order book of four stocks, namely Cisco (CSCO), Dell (DELL), Microsoft (MSFT) and Worldcom (WCOM). Before doing any analysis, we want to warn the reader that almost all results of our study may be affected by the fact that only the 15th highest bids and lowest asks orders are displayed at a given time. We checked that most of the activity takes place within the range of the 15th best orders, and we managed to keep track of consistently more than 75\% of all orders, but the reconstructive method we designed distorts the reality to an extend that is hard to estimate. Finally, note that we may not pick up an order whose size is greater than 10000 shares since it does not need to be displayed, even if it exists in the market. 

\section{Static properties}
 An order is anonymous, i.e. there is no way to identify its owner, but each order has a serial number, allowing us to study the fate of every order. 

\subsection{Order size}
\begin{figure}
\centerline{\psfig{file=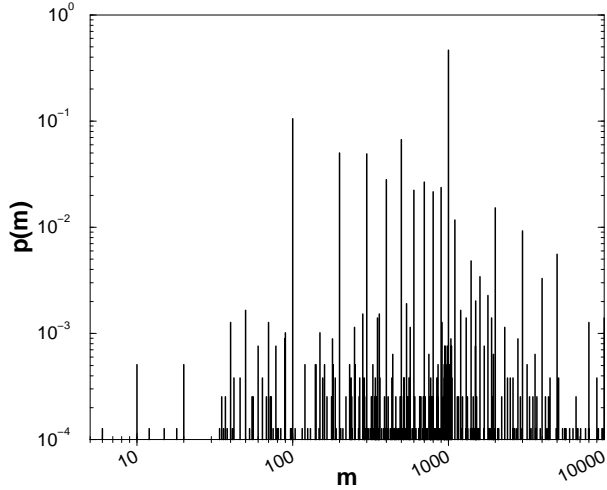,width=8cm}}
\caption{Order size probability distribution $P(m)$ of the bid size of DELL (10.04.2001). The clustering is obvious.}
\label{indsize}
\end{figure}

It is well known that orders have a tendency to cluster both in size and position \cite{Eco}: they tend to have a size of multiple of 10, 100 or 1000, and to be placed at  round prices, or at halves.  Fig \ref{indsize} reports $P(m)$, the probability distribution of their size for various days of CSCO, and clearly shows this clustering. We found also that the density of selected sizes does not depend on the stock nor on the day.

\subsection{Order lifetime}

\begin{figure}
\centerline{\psfig{file=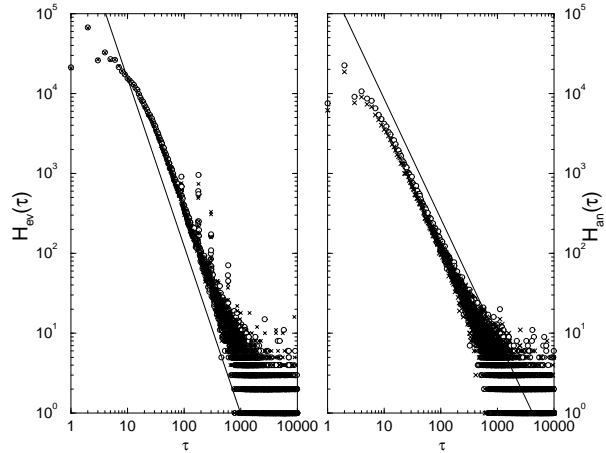,width=8cm}}
\caption{Lifetime histogram for bid (x) and ask orders (circles) of all orders of all stocks and all days. The left graph plots  $H_{\rm ev}$, the histogram for cancelled orders (straightline: power-law with a $-2.1$ exponent). The right graph plots $H_{\rm an}$, the histogram for fully filled orders (straightline: power-law with a $-1.5$ exponent).}
\label{lifetimeall}
\end{figure}

Fig \ref{lifetimeall} reports the histograms of the lifetime of all orders in our data. The lifetime distribution for cancelled and fully filled orders seem to have an algebraical decay with exponents that could\footnote{Our data actually only provide only an estimation of  the lifetime of each order, hence the value of these exponents are only indicative.} be $-2.1\pm0.1$ and $-1.5\pm0.1$ The lifetime distribution of cancelled orders has visible peaks at $\tau=90$, $180$, $300$ and $600$ seconds\footnote{The nature of our data broadens these peaks.}; it is clear that they correspond to typical timeouts. It is very likely that another typical timeout is $60$ seconds, although this peak does not clearly appear on Fig. \ref{lifetimeall}.

\subsection{Distribution shapes}
\label{shapes}
\begin{figure}
\centerline{\psfig{file=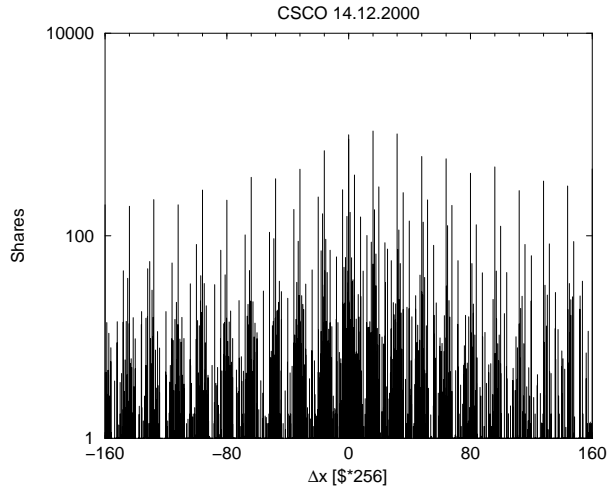,width=8cm}}
\caption{Histogram of aggregated orders for CSCO on 14.12.2000, showing that the orders are separated typically by \$~1/16  although the tick is \$~1/256.}
\label{priceclust}
\end{figure}

\begin{figure}
\centerline{\psfig{file=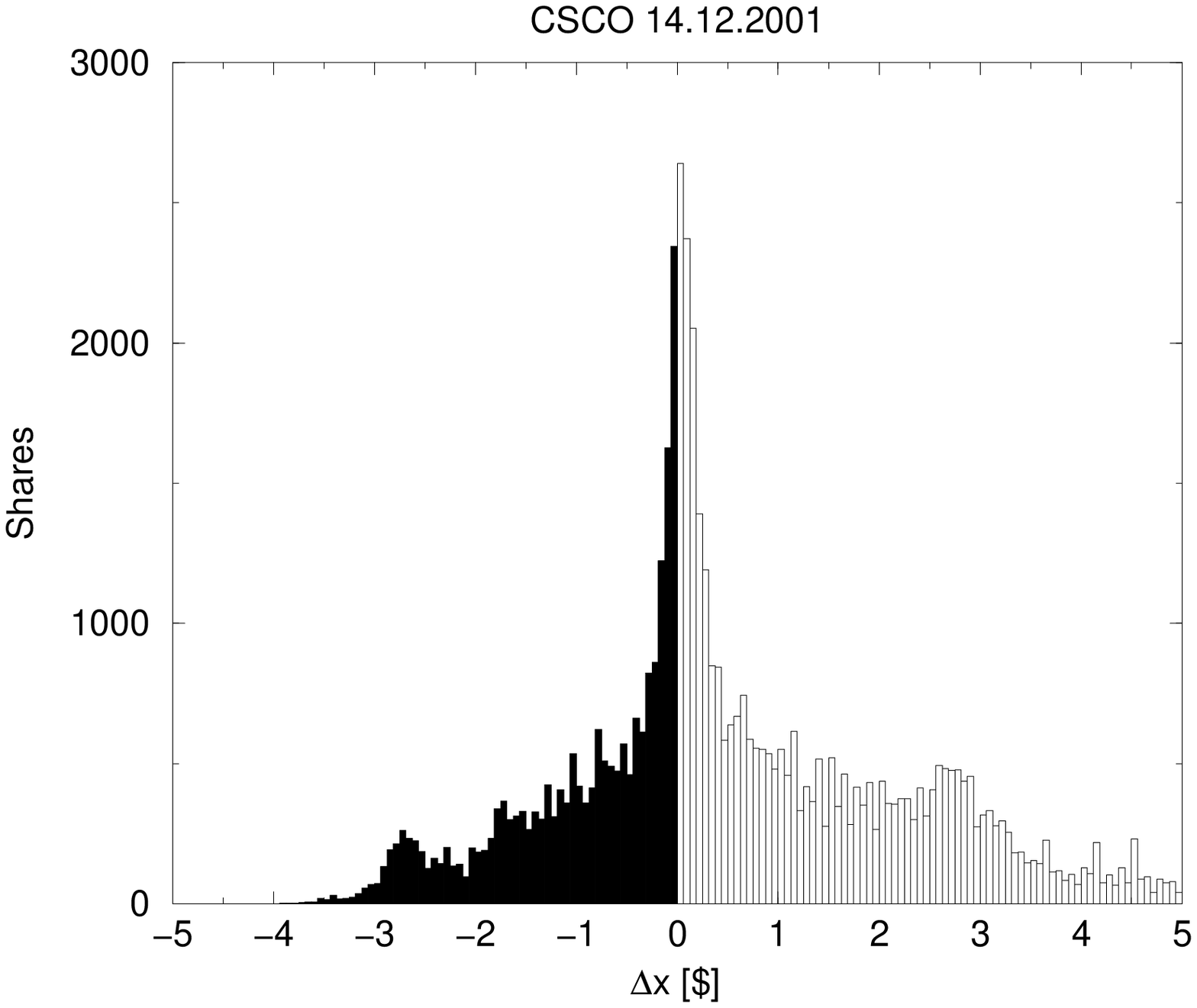,width=7cm}\psfig{file=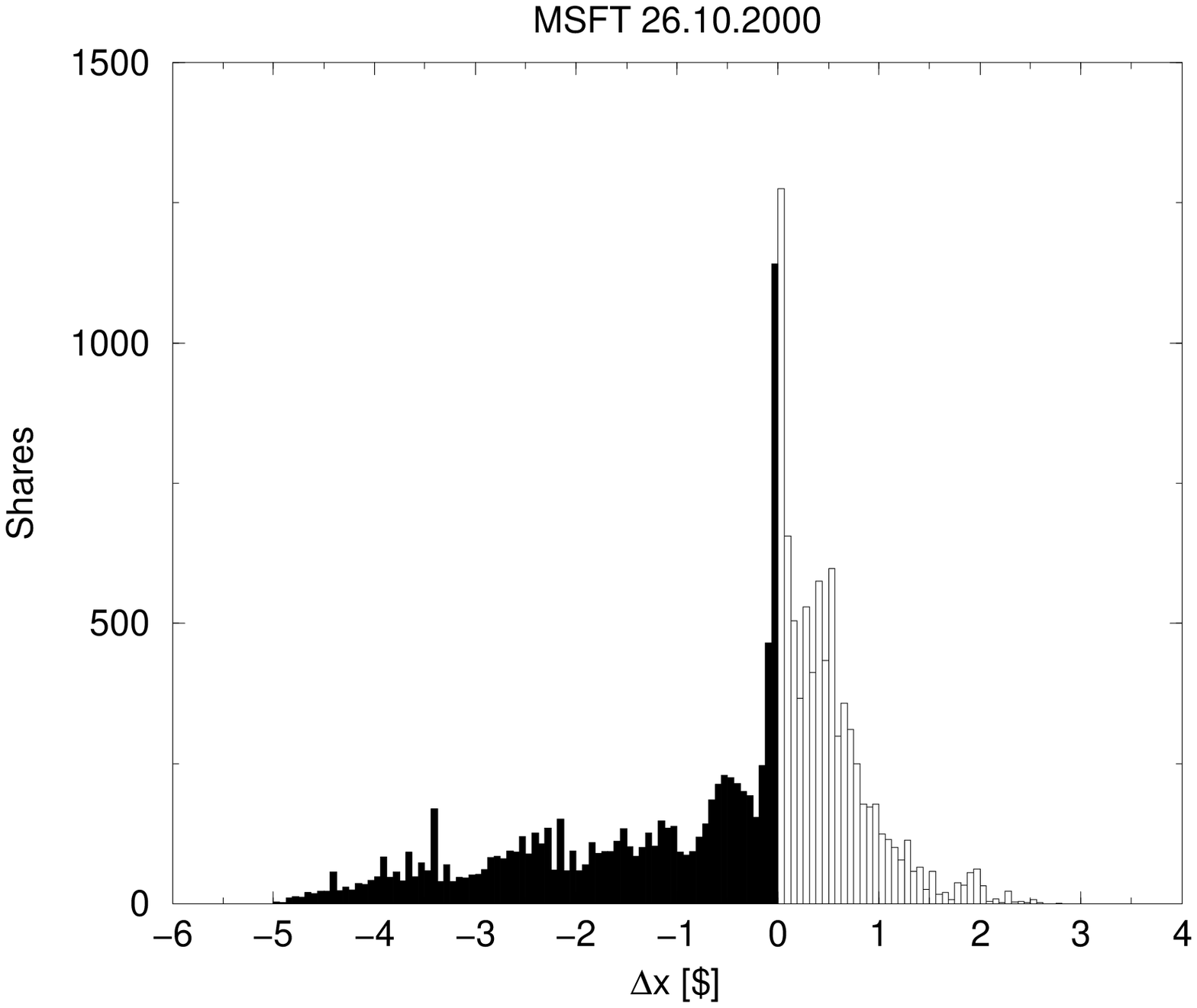,width=7cm}}
\caption{Histogram of aggregated orders of CSCO on 14.12.2000 (left graph), and MSFT on 26.10.2000 (right graph). The left distribution is the buy side and the right one the sell side.}
\label{sizerel}
\end{figure}

The shape of the orders distribution is a dynamical quantity in essence and can change tremendously in a few seconds. Plotting time-averaged distributions reveals some general features. 
For all time $t$, we define $A_t(p)$, and $B_t(p)$, which are the number of shares to be sold, and bought, at price $p$. Fig \ref{priceclust} plots the functions $a(\Delta x)=\avg{A_t(\Delta x_{\rm{ask}})}$ and $b(p)=\avg{B_t(\Delta x_{\rm{bid}})}$, where $\Delta x$ is the price relative to the relevant extremum of the considered distribution, that is, $\Delta x_{\rm{ask}}(t)=p(t)-a_m(t)\ge 0$ where $a_m$ is the price of the minimum ask, and $x_{\rm{bid}}(t)=p(t)-b_M(t)\le 0$, where $b_M$ is the price of the maximum bid. Note that the so-called {\em bid-ask spread} is then simply $g(t)=a_m(t)-b_M(t)$ The clustering in the position of the orders clearly appears on fig \ref{priceclust}: although the tick is \$~1/256, there are clear peaks separated by 16 ticks --- that makes \$~1/16, which is the official tick of the NASDAQ. Consequently, the effective tick is \$~1/16, and one has to coarse grain the raw distributions $a(\Delta x)\to a'(\Delta x)$ and $b(\Delta x)\to b'(\Delta x)$ by a factor 16; this is done on fig. \ref{sizerel}.

It appears that the distributions are convex and peaked at $\Delta x=0$.
This could explain the origin the overdiffusive behavior of prices which has been observed in various financial markets \cite{Bouchaud,MantegnaStanley}. Indeed, suppose that the order distributions are fixed during a time window and that the price follows a random walk, then the convexity of the orders distributions causes an over-diffusive price behavior.

 We could not find a general mathematical shape for distributions $a'(\Delta x)$ and $b'(\Delta x)$, maybe due to the fact that the data we have is not reliable for  the  tails of the distributions; we could note however that the buy and sell distributions are usually not symmetric, and this shows up through the virtual impact function discussed  in the next subsection.
 
\subsection{Market Impact}

\begin{figure}
\centerline{\psfig{file=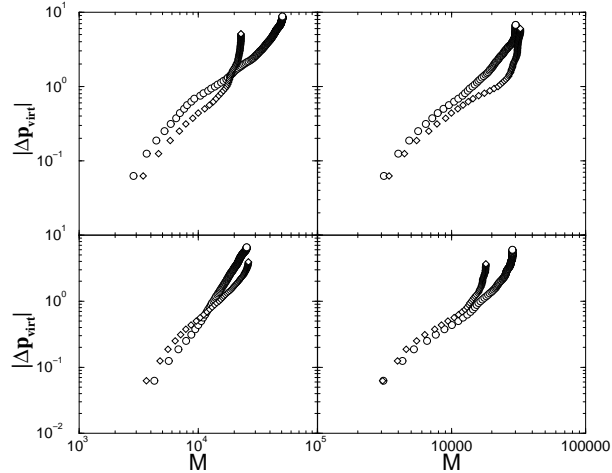,width=8cm}}
\caption{Virtual impact of four days of CSCO. Circles correspond to the ask side, and diamonds to the bid size}
\label{virtimp}
\end{figure}

Market impact is an important quantity: it is defined as the price change $\Delta p$ caused by a large order\footnote{Usually, large orders are split into several smaller orders in order to reduce the market impact.} of size $M$, hence, it is a functional relationship between the size and the price change. There are actually two types of impact. The first one is the {\em virtual impact}, which is related to the integral of the order distributions: $\Delta p(M)$ is defined through $M=|\int_0^{\Delta p} r(x)\ {\rm d}x|$, where $r(x)=a(x)$ or $b(x)$. Since apparently the average shape of the distributions have no general characteristic form, one can expect that the virtual market impact function differs from stock to stock and from day to day. This is the case, as shown in fig. \ref{virtimp}.  In the literature, it has been argued \cite{ZMEM} and suggested from data \cite{MaslovMills} that this impact should have a power-law form, with an 0.5 exponent \cite{ZMEM}, or with an exponent near 2 \cite{MaslovMills}. From our data, which are not reliable for the distribution tails, we could infer that the virtual impact functional form may not be strictly speaking a power-law, but can be approximated by such a form, with an exponent ranging from 1 to 3, depending on the day and stock. On the other hand, the {\em real impact} is defined as the real price change for a real large order, or a fast succession of almost simultaneous orders; the two types of impact can significantly differ.

\section{Dynamical properties}

Dynamical analysis\footnote{The analysis presented here can be compared with that in ref \cite{Coppejans}.} is made easier by a straightforward analogy between orders and particles:
an order can be thought as a particle whose mass is the order's size, and whose price is a spatial position. Since the possible prices are discretized, it makes sense to consider particles on a lattice whose mesh size is equal to the real tick. Placing an order is equivalent to {\em depositing} a particle, withdrawing an order amounts to {\em evaporating} it. A transaction takes places whenever two particles of opposite type are deposited at the same price, most often sequentially: this can be thought as a {\em annihilation}; note that an annihilation is caused by a deposition, hence there are two types of deposition depending on where an order is placed.

This analogy is already used for the modelling of limit orders markets in \cite{BPS,kogan}, where two types of particles-orders diffuse on a lattice and annihilate whenever they meet, and in \cite{Maslov}, where orders do not diffuse, but are randomly deposited and annihilated. Hereafter, we consider sell (ask) orders as type $A$ particles, and buy (bid) orders as type $B$ particles. The terminology used hereafter is that of physics; in particular, we will use the word gap instead of spread.

\subsection{Temporal properties; rates}
\label{autocorr}

We consider three types of event, namely deposition, evaporation and annihilation\footnote{An annihilation is always caused by a deposition of an opposite particle, but we do not include this type of event into the deposition rates.}, and denote their rate by $\delta$, $\eta$ and $\alpha$ respectively.
Whenever needed, subscript $B$ labels rates associated with $B$-type particles, and $A$ with $A$-type particles. A rate is defined as the average number of events per second. 

\begin{figure}
\centerline{\psfig{file=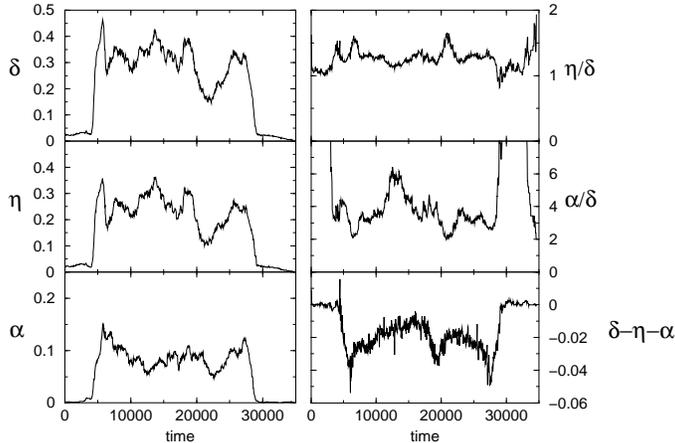,width=8cm}}
\caption{On the left, from top to bottom: bid rates of deposition $\delta$, evaporation $\eta$ and annihilation $\alpha$ in events/second, time averaged over a window of 1500 seconds of a day of DELL. Right: ratios  $\delta/\eta$ and $\delta/\alpha$, and $\delta-\eta-\alpha$}
\label{rates}
\end{figure}
\begin{figure}
\centerline{\psfig{file=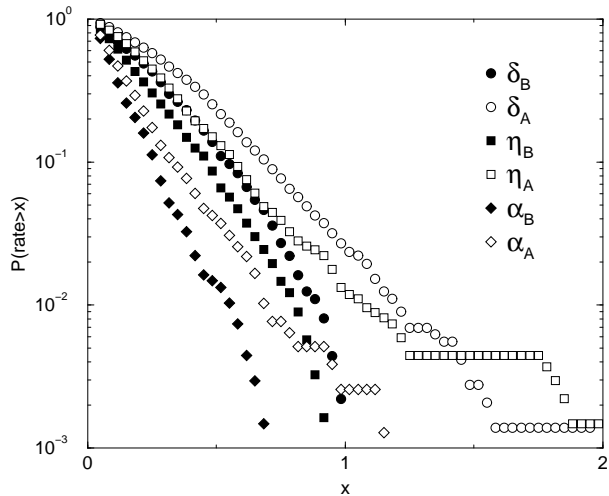,width=8cm}}
\caption{Cumulative probability of event rates $\delta$ (circles), $\eta$ (squares)  and $\alpha$ (diamonds) of  bid (filled symbols) and ask (plain symbols) sides (DELL 09.04.2001) ($Dt=30$~s).}
\label{prates}
\end{figure}
\begin{figure}
\centerline{\psfig{file=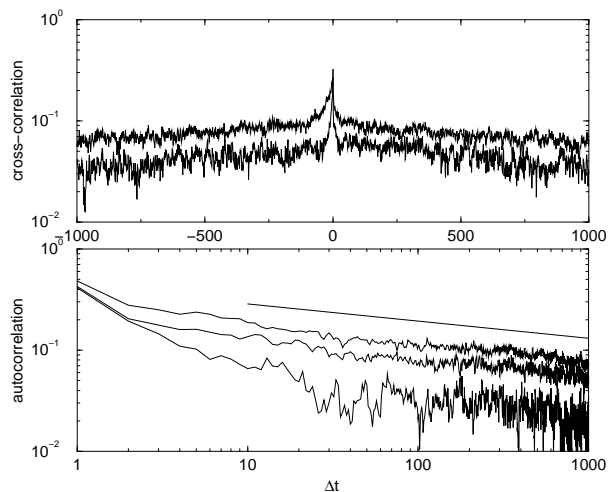,width=8cm}}
\caption{Bottom graph: autocorrelation of the bid rates $\delta$ (top), $\eta$ (middle) and $\alpha$ (bottom) measured with $\Delta t=1$ s the continuous line has a -0.17 exponent. Upper graph: cross-correlations $\avg{\delta(t)\eta(t+\Delta t)}$ (top) and $\avg{\delta(t)\alpha(t+\Delta t)}$ (DELL 09.04.2001)}
\label{ratescorr}
\end{figure}

The method we used to plot fig \ref{rates} was to count how many events occurred during a given time window, typically $Dt=30$ seconds, then of course to divide this number by $Dt$, and to run a moving average in order to smooth the graphs. Figure \ref{rates} reports the $B$ rates of one day of DELL. They fluctuate wildly during the day, but not in an independent way: the autocorrelation and crosscorrelation functions of $\delta$, $\eta$ and $\alpha$ are slowly decreasing, apparently algebraically, as shown by fig. \ref{ratescorr} for $\delta$, $\eta$ and $\alpha$; the gap was also found to share this property. It is tempting to consider these facts as the origin of the well-known algebraical decay of the autocorrelation of the price volatility. Note that $\avg{\delta(t)\eta(t-\Delta t)}>\avg{\delta(t)\eta(t+\Delta t)}$ ($\Delta t>0$), which is due in part to the redeposition of evaporated orders (see also  section  \ref{diffus}). The distribution of the number $n(t)$ of events during $t$ and $\Delta t$ has exponential, hence Poissonian tails  (fig. \ref{prates}), but this distribution is computed over a whole day, and is therefore a superposition of a time-varying distribution. The fact that a Poissonian distribution is obtained is merely a consequence of this superposition and the associated loss of information.

We emphasize that the evaporation rate $\eta$ is high, which is consistent with the short typical order lifetimes and the fact that removing an order from Island ECN is free of charge\footnote{Hence the typical evaporation rate may be smaller in ECNs which charge for cancelling orders.}. We found that typically 80\% of all orders evaporate, and 20\% anihilate, in agreement with results of ref \cite{Coppejans}, where similar evaporation rates were also measured in the OM Stockholm and London Securities and Derivative Exchange . This implies that the absence of evaporation in model of ref. \cite{Maslov} is an important shortcoming.

There are some robust properties of the dynamics of the books: the right column of figs. \ref{rates} shows that the ratio $\eta/\delta$  is quite constant during the usual trading period. On the other hand, the ratio $\alpha/\delta$ is obviously related to the real trading activity and shows periods where it remains constant, but can greatly vary during a day, suggestting that the composition of the market's population has changed. The way the rate $\alpha$ is defined implies that an annihilation does not implies a loss of a particle (orders can be partly filled), hence  $\eta+\alpha$ is consistently greater than $\delta$ and the particle conservation law should read $\eta/+\alpha'=\delta$, where $\alpha'/\alpha={\rm cst}$  . On the other hand, the mass rates $\delta^v$, $\eta^v$ and $\alpha^v$, which count how many shares deposit, evaporate and annihilate per second, often obey a mass conservation relationship $\delta^v=\eta^v+\alpha^v$ during a trading day; we checked that these rates also have algebraically decreasing auto- and cross-correlation functions. 

\subsection{Spatial properties}

\begin{figure}
\centerline{\psfig{file=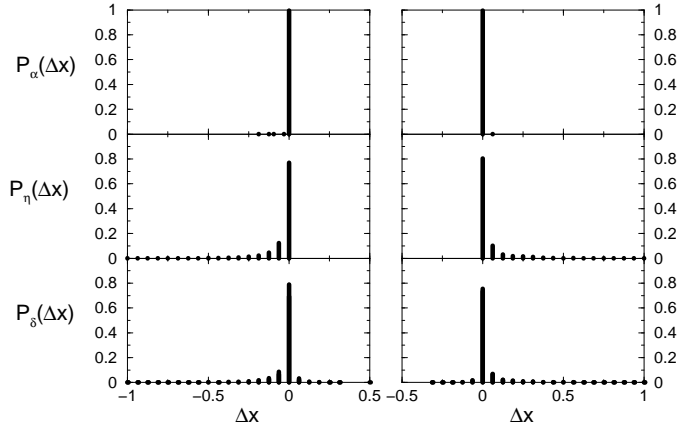,width=8cm}}
\caption{Frequency of an event versus its relative position (DELL 09.04.2001) (left column: bid side; right column: ask side)}
\label{spatial_distr}
\end{figure}

It is known that the activity in order driven markets occur near the relevant extrema of the distributions \cite{Eco,Hasbrouck}. The shape of the orders distribution (see section \ref{shapes}) is somewhat a confirmation of this statement, but the rate of activity is an even better one. Fig. \ref{spatial_distr} reports the frequency of the three fundamental events versus their relative position $\Delta x$. It is clear from that figure that all events usually occur close to $\Delta x=0$. By comparing the probability of an event at position $\Delta x$ (fig. \ref{spatial_distr}) with the average number of shares at that position (fig. \ref{sizerel}), which is in first approximation proportional to the number of orders, one can see that the probability of an event depends on $x$.

\subsection{Spatio-temporal properties}

\begin{figure}
\centerline{\psfig{file=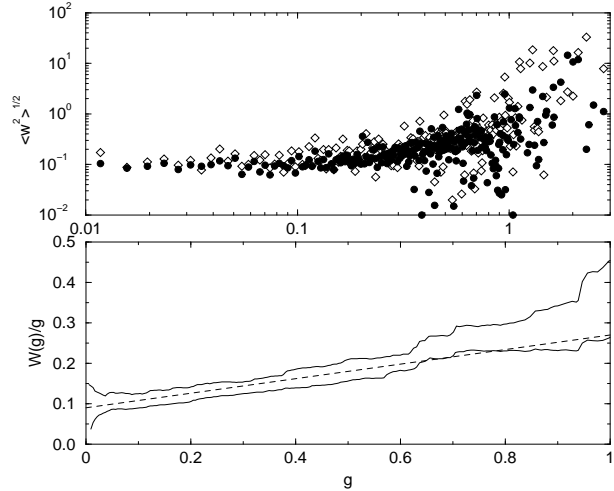,width=8cm}}
\caption{Upper graph: spatial width of deposition versus gap averaged over all stocks and all days (bid: . Lower graph: normalized integral $W(g)/g$ of $w(g)$ (bids: thick curve, asks: thin curve), and a reference line $y=0.18~x+C$ (dotted line).}
\label{ALL-width}
\end{figure}

This type of property is much harder to analyse. Nonetheless, it appears that the deposition process depends on the gap $g$. Indeed, defining the spatial width of deposition $w=\sqrt{\avg{\Delta x_{\rm dep}^2}}$, one obtains fig \ref{ALL-width} which reports $w_A(g)$ and $w_B(g)$. We find that $w$ is constant up to $g\simeq \$\ 1/16$,\footnote{this is the real basic length scale of this system (see above).} i.e. as long as the two order distributions are adjacent; when $g\ge \$\ 1/8$, i.e. when there is an effective tick between the two distributions, we found a {\em linear relationship} between $w$ and $g$ for small values of $g$, as shown by $W(g)/g=\int_0^g w(g'){\rm d}g'/g$, the normalized integration of $w(g)$, which suppresses the noise. 

We found a linear relationship for both orders that are deposited onto the gap, and the ones that are placed in the bulk of the distributions. It is easy to see how gap orders lead to this relationship: suppose that some trader is willing to pay less than the current best offer, but still in a hurry, then she will place an order anywhere in the gap, e.g. at the middle point, causing the observed effect. On the other hand, it could be possible that the linear dependence measured for bulk deposition were due to the small price window defined by the 15 best orders on each side; however, we found the this window is typically of order \$~0.8, whereas fig \ref{spatial_distr} shows that most of the activity occurs in a small partof  the price window. This is an evidence that allows us to conclude that this linear dependence is genuine.

\subsection{Diffusion}
\label{diffus}
\begin{figure}
\centerline{\psfig{file=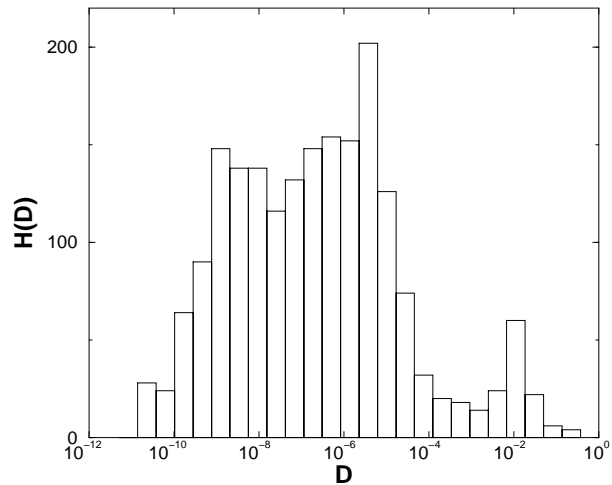,width=8cm}}
\caption{Histogram $H(D)$ of individual order diffusion coefficients. The vast majority of orders diffuse less than one effective tick during their life.}
\label{Diff}
\end{figure}

We could also test if orders diffuse on the lattice, i.e. if an order jumps from a price to another before meeting an order of opposite type and annihilating, or simply evaporates. Orders diffusion is the central assumption of models found in ref \cite{BPS,kogan}. The method we used is to keep track of orders with unusual sizes, for instance 832 shares. When such an order first evaporates, we begin to record its trajectory, until it is annihilated. We found that about 2/3  of such orders are never deposited again, i.e. they simply vanish; if they are deposited again, their average number of redeposition is about 2; amongst the redeposited orders, about half of them are redeposited at the same price a few seconds later, which shows up in the asymmetry of the cross-correlation function $\avg{\delta(t)\eta(t-\Delta t)}$ (fig \ref{ratescorr}). The diffusion coefficient $D$ of the diffusive orders turns out to be typically $D\le10^{-5} [\$^2/s]$ (see Fig \ref{Diff}), which corresponds to an average price deviation of one effective tick during $1000$s \footnote{Our method actually sometimes overestimates the diffusion coefficient, since it may happen that a odd-sized order that evaporates belongs to another person when it is deposited some time later.}. This suggests that the fundamental process of the orders dynamics is not diffusion; however such an approach is useful, particularly concerning analytical results \cite{BPS,Tang,kogan}.

\section{A new breed of models for limit orders markets}

We believe that the above facts have strong consequences for the modelling of limit order markets. They imply that one has to reconsider the two existing families of non-strategic models of limit orders, those of Bak, Paczuski and Shubik (BPS) \cite{BPS,kogan}, the one of  Maslov \cite{Maslov}. Both are very appealing thanks to their extreme simplicity, which makes them suitable for analytical analysis\footnote{See ref \cite{Tang} for an analytical approach of BPS models, and \cite{Slanina} for a mean-field solution of Maslov's model.} and they are able to mimick some typical market behavior, such as volatility clustering and large price variation probability density function, but both include fundamental hypotheses that are in part incompatible with our findings. In addition, at least under their simplest form, they have a price behavior that differs from what is observed in real markets. Indeed,  the returns $r_{\Delta t}(t)$ between time $t-\Delta t$ and $t$, usually defined as $\log p(t)-\log p(t-\Delta t)$, where $p(t)$ is the price at time $t$,  are known to be overdiffusive for short times, i.e.  $\avg{r_{\Delta t}}_t\propto t^{H(\Delta t)}$ with $H(\Delta t)>1/2$ for small $\Delta t$, and to tend to those of a random walk for large times, i.e. $H(\Delta t)\simeq 1/2$ for $\Delta t\propto $ a year \cite{Bouchaud}. The basic BPS model and Maslov's model are characterized by $H(\Delta t)=1/4$ for all $\Delta t$, hence the prices are underdiffusive and do not tend to a random walk in the long time limit. However, as soon as the orders diffusion is biased towards the gap in BPS model,  $H=1/2$ is recovered for large times $\Delta t$. Obtaining an over-diffusive behavior seems to require an (strategic) imitating behavior that leads to some kind of herding \cite{BPS}.

Our aim is to build a model compatible with our observations. We retain two processes, namely deposition and evaporation: an annihilation being first caused by the deposition of a new order on top of a previously existing opposite order, it can be included into the deposition process. In addition, we consider that all positions are relative to the relevant extremum of the order distributions, i.e. $a_m(t)$ and $b_M(t)$\footnote{This contrasts with Maslov's model, where events occur relative to the last paid price, but may not be crucial.}. For the time being, we restrict the mass of orders to be all the same, and consider all events to be independent; we also suppose that only one order can be deposited during a timestep, that the rates are constant, and shall discuss the consequence of each hypothesis in the following.

\subsection{Definition of the model}

We consider an infinite  one dimensional lattice and two types of particles, $A$ and $B$ corresponding to the types of order. These can be  considered as particular states of $X-$particles ($X\in\{A,B\}$); two types of events can happen\footnote{All quantities below can of course depend on the type of particle.}:

\bit
\item Deposition: with probability $\delta$, a $X-$particle is deposited at site $x$ drawn at random from the pdf $P_X(x,t)$. Whenever the particle $X$ is deposited into the bulk of the $-X$ distribution, one considers that the annihilation takes place at the relevant extremum of the $-X$ distribution
\item Evaporation: each $X-$particle on the lattice has a probability $\eta$ to evaporate.
\eit

In order to complete the definition of the model, the pdf $P_X(x,t)$ has to be specified. For the sake of simplicity, we consider a Gaussian function.. For the time being, we assume that it is centered on the relevant extremum of the particle distribution. Its variance $\sigma_X(t)$ can either be constant, or linearly depend on the gap: $\sigma_X(t)=K g(t)+C$; if $C=0$, $K$ fixes the annihilation rate $\alpha$, and if $C\ne 0$, $\alpha$ increases as the gap decreases. The fact that every particle has the same probability of evaporation implies that the number of particles has a well defined average and may be seen as equivalent to a time-out of order $1/\eta$. This model has therefore four parameters, $\delta$, $\eta$, $K$ and $C$. 

\subsection{Results}

\begin{figure}
\centerline{\psfig{file=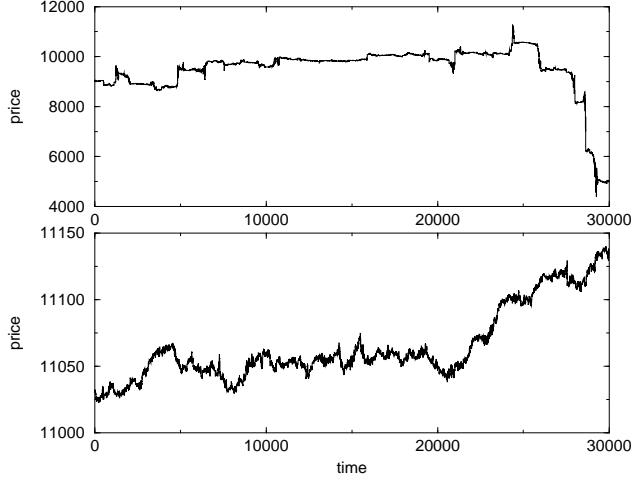,width=8cm}}
\caption{Two time series of the market model with parameters $\alpha=1/2$, $\delta=1/100$, $K=2$, $C=3$ (upper graph) and $K=1$ (lower graph), showing the dramatic influence of $K$ on the system's properties.}
\label{timeseries}
\end{figure}

\begin{figure}
\centerline{\psfig{file=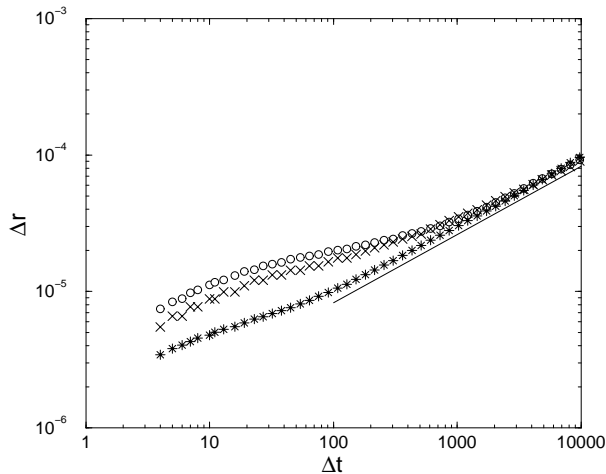,width=8cm}}
\caption{Results from the market model: typical $\avg{r_{\Delta t}}$ versus $\Delta t$, showing that the Hurst exponent $H$ is 1/4 for small times and tends to 1/2 for large times (circles: at most one particle is deposited during $t$ and $t+1$ $\delta=0.5$, $\eta=0.01$, $K=1.5$, $C=10$; x: simultaneous deposition of an average of $\delta$ particles of both types per unit of time (same parameters); $\star$: simultaneous deposition $\delta=2$, $\eta=0.04$, $K=1.5$, $C=10$.}
\label{hurst}
\end{figure}

 We observed power tails of the returns and, depending on parameters, volatility clustering. The parameters have strong effects on the dynamics of the model (see figure \ref{timeseries}). As soon as evaporation takes place, the returns have a random walk behavior at large times (fig. \ref{hurst}).  This means that evaporation is probably enough to remedy the underdiffusive behavior of prices even at large times of Maslov's model.

One of the main assumptions of the model we propose is that at most one order is deposited during a timestep. This is in clear contradiction with markets data. Let us extend our model in order to include simultaneous depositions. If at each time step, the number $n_X(t)$ of deposited orders of type $X$ is drawn from an exponential distribution with average $\delta$, the region where the price is under-diffusive becomes gradually less pronounced when $\delta$ is increased, even while keeping $\delta/\eta$ constant. For the time being, the present model has no overdiffusive behavior, which seems to be a property shared by all simple market particles models with no strategic behavior. The effect of the non-trivial time properties of the real rates is still unclear, but being the consequence of agents behavior, it may be an indirect origin  of the overdiffusive behavior of financial markets. In addition, we note that in our model as well as in Maslov's model, the presence of volatility clustering is due to the algebraical decay of the gap autocorrelation function. Finally, the effect of heterogeneity in order size is still to be determined.

\section{Conclusions}

We have determined several features of  Island ECN's order books. Some of our findings may be specific to this ECN. The static features we studied include the size and life-time probability distributions, the average shape of the orders distribution, from which we derived the several market impact functions. More importantly, we have been able to determine fundamental mechanisms of such markets, which has important consequences for the modelling: this lead us to consider alternative over-simple particles models, which offer a closer description of real markets. We have observed that the under-diffusive behavior of simple particles models is less pronounced if the deposition of more than one particle per time step is allowed   to deposit. Finally, a non-strategic behavior that leads to overdiffusive price behaviors in particles models has still to be found, and may require non-constant rates with non-trivial interdependence.

We thank Y.-C. Zhang, S. Maslov, M. Marsili, and Th. Bochud for useful discussions. This work has been supported in part by the Swiss National Funds of Scientific Research.

\end{document}